\documentclass[twocolumn,showpacs,superscriptaddress,preprintnumbers,prd]{revtex4}
\usepackage{graphicx,amsmath,dcolumn,bm}
\begin{document}
\preprint{KEK-TH-1167}
\title{\Large \bf Proposal for exotic-hadron search by fragmentation functions}
\author{M. Hirai}
\affiliation{Department of Physics, H-27,
             Tokyo Institute of Technology,
             Meguro, Tokyo, 152-8551, Japan}
\author{S. Kumano}
\affiliation{Institute of Particle and Nuclear Studies,
          High Energy Accelerator Research Organization (KEK) \\
          1-1, Ooho, Tsukuba, Ibaraki, 305-0801, Japan}
\affiliation{Department of Particle and Nuclear Studies,
             Graduate University for Advanced Studies \\
           1-1, Ooho, Tsukuba, Ibaraki, 305-0801, Japan}     
\author{M. Oka}
\affiliation{Department of Physics, H-27,
             Tokyo Institute of Technology,
             Meguro, Tokyo, 152-8551, Japan}
\author{K. Sudoh}
\affiliation{Institute of Particle and Nuclear Studies,
          High Energy Accelerator Research Organization (KEK) \\
          1-1, Ooho, Tsukuba, Ibaraki, 305-0801, Japan}
\date{November 16, 2007}
\begin{abstract}
It is proposed that fragmentation functions should be used to identify
exotic hadrons. As an example, fragmentation functions of the scalar meson
$f_0(980)$ are investigated. It is pointed out that the second moments
and functional forms of the $u$- and $s$-quark fragmentation functions
can distinguish the tetraquark structure from $q\bar q$. By the global
analysis of $f_0 (980)$ production data in electron-positron annihilation,
its fragmentation functions and their uncertainties are determined.
It is found that the current available data are not sufficient to
determine its internal structure, while precise data in future should
be able to identify exotic quark configurations.
\end{abstract}


\maketitle


In the hadron mass region below 1 GeV, there are scalar mesons,
$f_0(600)$, $f_0(980)$, and $a_0(980)$, whose internal configurations
are not obvious \cite{summary}. Their flavor compositions could be
$f_0(600) = (u\bar u+d\bar d)/\sqrt{2}$,
$f_0(980) = s\bar s$,
$a_0(980) =u\bar d$, $(u\bar u-d\bar d)/\sqrt{2}$, $\bar ud$
in a simple quark model by considering the mass relation,
$m_u  \sim m_d < m_s$.
However, this assignment implies a mass sequence,
$m(f_0(600)) \sim m(a_0(980)) < m(f_0(980))$, which contradicts
with the observed one,
$m(f_0(600)) < m(a_0(980)) \sim m(f_0(980))$.
If $f_0(980)$ and $a_0(980)$ are exotic states such as tetraquark ones,
the observed spectrum could be naturally understood.
Since $f_0(980)$ and $a_0(980)$ are experimentally established resonances,
they provide a good opportunity to study exotic mesons beyond
a naive $q\bar q$-type quark model. 


First, a brief outline of recent studies is given for the $f_0(980)$
structure. In a simple quark model, a light scalar meson $f_0$ with
$J^{PC}=0^{++}$ is identified as a $^3 P_0$ quarkonium with the flavor
structure $(u\bar u+d\bar d)/\sqrt{2}$. However, if such an ordinary
$q\bar q$ configuration is assigned for $f_0(980)$, the strong decay width
is very large, $\Gamma(f_0 \rightarrow \pi\pi)=500-1000 \ {\rm MeV}$,
according to various theoretical calculations \cite{f0width-th}.
The small experimental width 40$-$100 MeV \cite{pdg} cannot be
consistently explained by simple quark models.

The strong-decay width suggests that $f_0(980)$ should not be an ordinary
nonstrange $q\bar q$-type meson. The Fermilab-E791 collaboration measured
the decay $D_s^+ \rightarrow \pi^- \pi^+ \pi^+$ \cite{e791},
which can proceed via intermediate states, for example,
$D_s^+ \rightarrow f_0 (980) \pi^+$ with $s\bar s$ quarks in $f_0 (980)$.
This experiment suggested a sizable strange-quark component in $f_0 (980)$.
The simplest configuration is a pure strange quarkonium $s\bar s$
for $f_0 (980)$. In addition, since its mass is just below the $K\bar K$
threshold, it could be considered as a $K\bar K$ molecule \cite{kkbar}.
If two color-singlet states of $K$ and $\bar K$ are not well separated,
it corresponds to a tetraquark state,
$(u\bar u s\bar s+d\bar d s\bar s)/\sqrt{2}$, which was originally
suggested in the MIT bag model \cite{tetra}. Recent QCD-sum-rule studies
support this idea of a tetraquark state \cite{qcdsum07}. Furthermore,
there are lattice-QCD studies that $f_0(980)$ corresponds to the tetraquark
state because the scalar tetraquark mass is about 1.1 GeV 
\cite{lattice-4q}. In addition, $f_0(980)$ used to be considered
as a glueball candidate; however, recent lattice QCD calculations rule out
such a possibility because the mass of a $0^{++}$ glueball is estimated
about 1700 MeV \cite{lattice-glue}. The situation of scalar mesons with
masses in the 1 GeV region is summarized in Ref. \cite{summary}.
All the possible $f_0(980)$ configurations are listed in Table
\ref{tab:f0-config} although the nonstrange-$q\bar q$
and glueball states seem to be unlikely according to the recent studies.

\begin{table*}[t]
\caption{Possible $f_0(980)$ configurations and their features 
         in fragmentation functions at small $Q^2$.}
\label{tab:f0-config}
\begin{ruledtabular}
\begin{tabular}{cccc}
Type                   & Configuration 
                       & Second moments
                       & Peak positions      \\
\colrule
Nonstrange $q\bar q$   & $(u\bar u+d\bar d)/\sqrt{2}$  
                       & $M_s<M_u<M_g$
                       & $z_u^{\rm max}>z_s^{\rm max}$   \\  
Strange    $q\bar q$   & $s\bar s$                 
                       & $M_u  <   M_s \lesssim M_g$    
                       & $z_u^{\rm max}<z_s^{\rm max}$   \\  
Tetraquark (or $K\bar K$) & $(u\bar u s\bar s+d\bar d s\bar s)/\sqrt{2}$  
                       & $M_u \sim M_s \lesssim M_g$
                       & $z_u^{\rm max} \sim z_s^{\rm max}$   \\  
Glueball               & $gg$ 
                       & $M_u \sim M_s < M_g$
                       & $z_u^{\rm max} \sim z_s^{\rm max}$   \\  
\end{tabular}
\end{ruledtabular}
\end{table*}
\vspace{+0.0cm}

In the following, the notation $f_0$ indicates the $f_0(980)$ meson 
and $f_0(600)$ is not discussed.
There were proposals to find the structure by a $\phi$ radiative
decay into $f_0$ \cite{cik93,ag97,phi1}. Since it is an electric dipole
decay, the width should reflect information on its size, namely its internal
structure \cite{cik93}. The experimental measurements of VEPP-2M \cite{vepp}
and DA$\Phi$NE \cite{daphne} were reported for the decay
$\phi \rightarrow f_0 \gamma$.
The data may suggest the tetraquark picture; however, there are still
discussions on their interpretation \cite{phi1}.
Another possible experimental probe is the $\gamma\gamma\rightarrow\pi^+\pi^-$
process in the $f_0$ mass region. The two-photon decay width of $f_0(980)$
was recently reported as
$0.205^{+0.095}_{-0.083}({\rm stat})^{+0.147}_{-0.117}({\rm syst})$ keV
by the Belle collaboration \cite{belle07}.
Model calculations indicate $1.3-1.8$ keV in the nonstrange 
$q\bar q$ picture; however, the measurement is consistent with
the $s\bar s$ and $K\bar K$-molecule configurations. 
There are also ideas to use elliptic flow and nuclear modification ratios
in heavy-ion reactions for finding exotic hadron structure \cite{heavy-ion}.

There are compelling theoretical and experimental evidences that
the scalar meson $f_0(980)$ is not an ordinary nonstrange $q\bar q$ meson.
However, a precise configuration is not determined yet, and
a clear experimental evidence is awaited.
It is the purpose of this paper to show that the internal structure of
exotic hadrons should be determined from their fragmentation functions 
by noting differences in favored and disfavored functions. We investigate
$f_0(980)$ as an example in this work.


A fragmentation function is defined by a hadron-production cross section
and the total hadronic cross section:
$
F^h(z,Q^2) = \frac{1}{\sigma_{tot}} 
\frac{d\sigma (e^+e^- \rightarrow hX)}{dz} .
$
Here, the variable $z$ is defined by the hadron energy $E_h$ and
the center-of-mass energy $\sqrt{s}$ ($=\sqrt{Q^2}$) by
$z \equiv E_h/(\sqrt{s}/2)$. The fragmentation occurs from primary partons,
so that it is expressed by the sum of their contributions:
$
F^h(z,Q^2) = \sum_i C_i(z,\alpha_s) \otimes D_i^h (z,Q^2),
$
where $\otimes$ indicates the convolution integral,
$f (z) \otimes g (z) = \int^{1}_{z} dy f(y) g(z/y)/y$,
$D_i^h(z,Q^2)$ is the fragmentation function
of the hadron $h$ from a parton $i$ ($=u,\ d,\ s,\ \cdot\cdot\cdot,\ g$),
$C_i(z,\alpha_s)$ is a coefficient function,
and $\alpha_s$ is the running coupling constant.
The favored fragmentation means a fragmentation from a quark
or an antiquark which exists in a hadron as a constituent
in a quark model, and the disfavored means a fragmentation
from a sea quark. The favored and disfavored functions are assigned
in the following discussions by considering the naive quark 
configurations in Table \ref{tab:f0-config}.

\begin{figure}[b]
\includegraphics[width=0.34\textwidth]{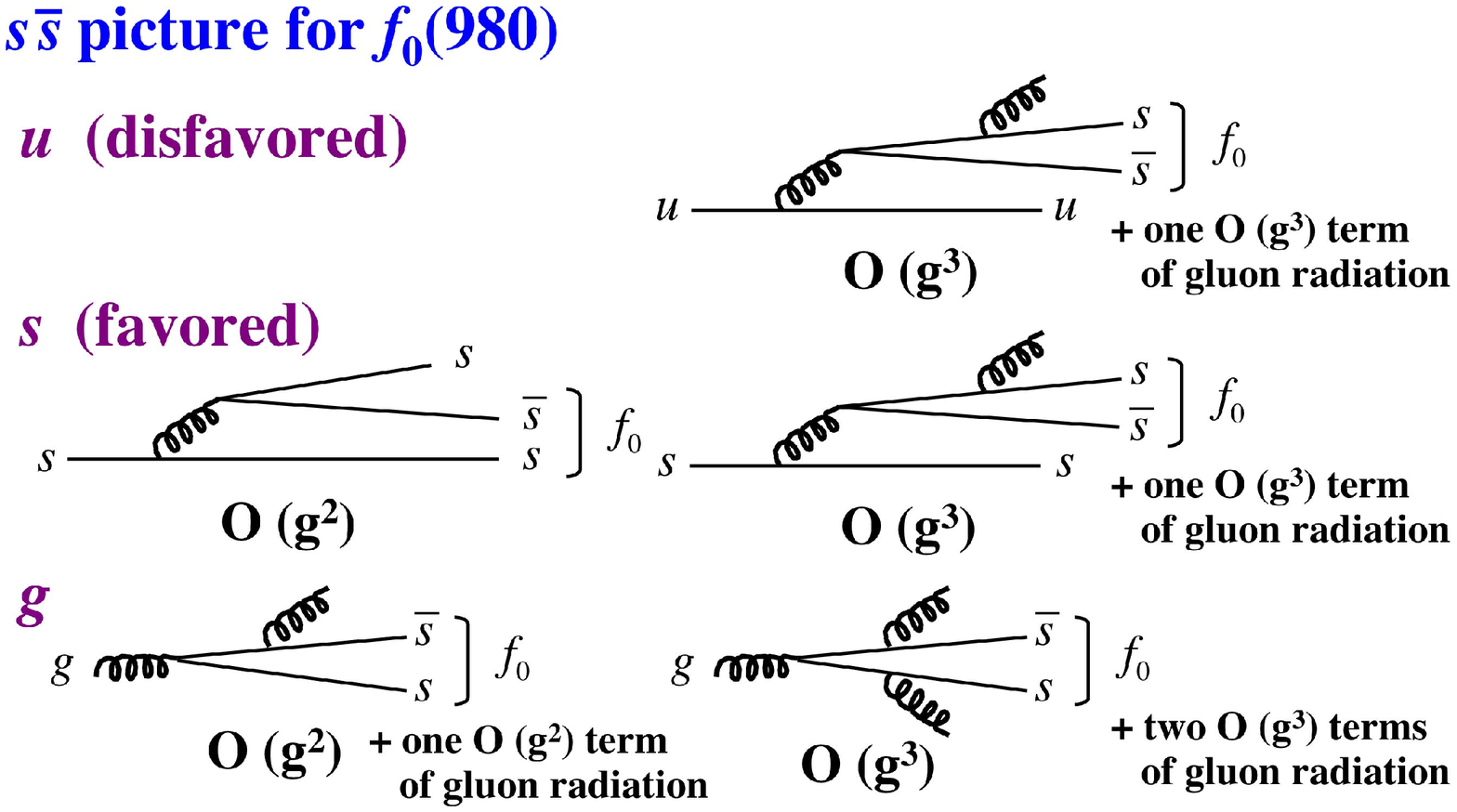}
\vspace{-0.3cm}
\caption{(Color online) Schematic diagrams for $f_0$ production
          in the $s\bar s$ configuration.}
\label{fig:ssbar}
\vspace{+0.5cm}
\includegraphics[width=0.34\textwidth]{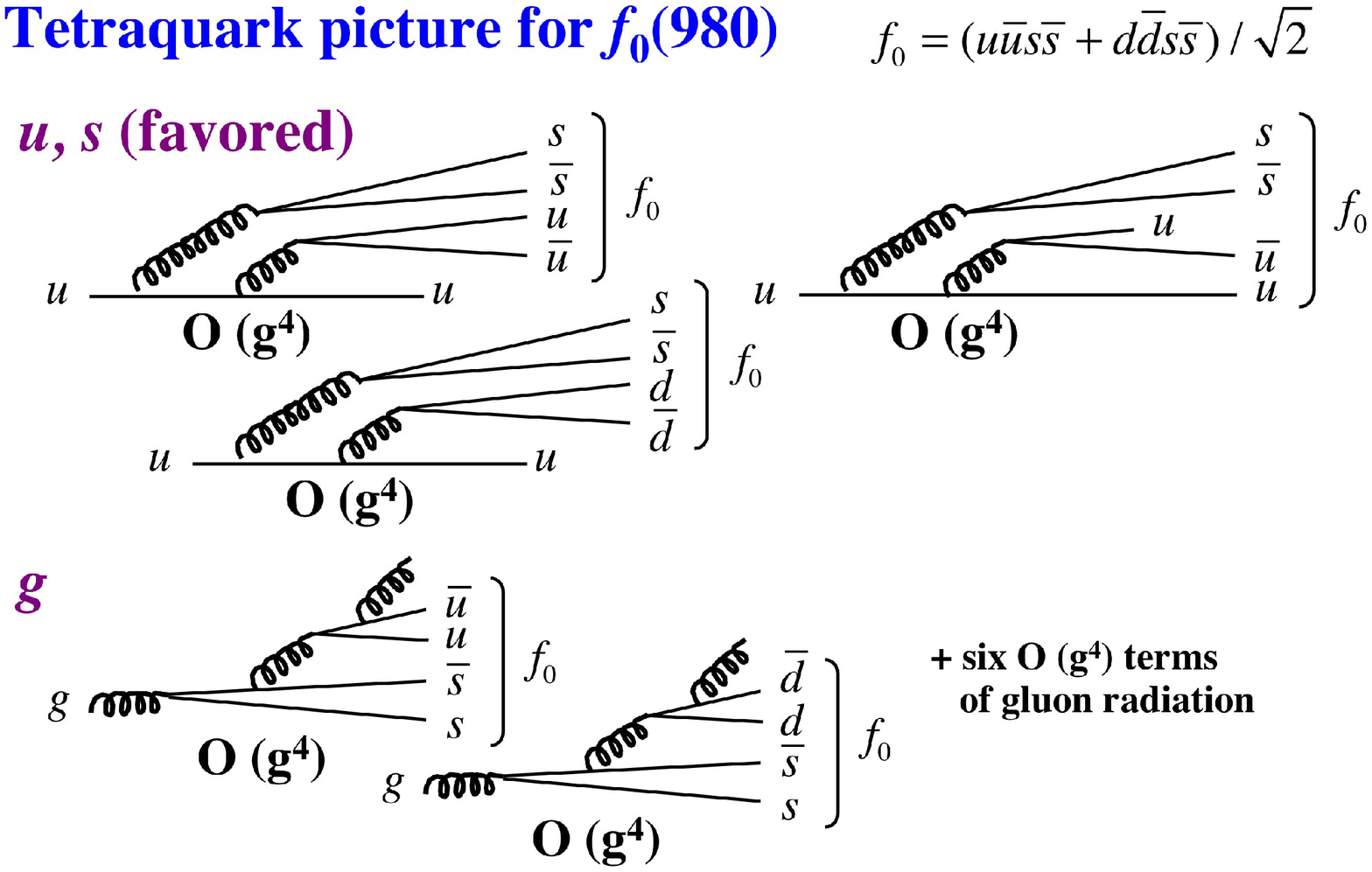}
\vspace{-0.3cm}
\caption{(Color online) Schematic diagrams for $f_0$ production
          in the tetraquark configuration.}
\label{fig:tetraquark}
\end{figure}

We first consider a possible $s\bar s$ configuration for $f_0$.
Then, the $u$- and $d$-quark fragmentation functions are disfavored
ones and the $s$-quark function is a favored one. For example,
the favored fragmentation from $s$ is possible if a gluon
is radiated from $s$, and then it splits into a $s\bar s$ pair
to form the $f_0$ meson as shown in Fig. \ref{fig:ssbar}.
The notations $O(g^2)$ and $O(g^3)$ indicate the second and third
orders of the coupling constant $g$.
In the disfavored process from $u$, there are processes in the order
of $O(g^3)$ without an $O(g^2)$ term, so that its probability is
expected to be smaller than the favored one from $s$. It leads to
the relation for the second moments of fragmentation functions:
$M_u < M_s$, where $M_i \equiv \int dz z D_i^{f_0} (z)$. 
The second moment $M_i$ is the energy fraction for $f_0$
which is created from the parton $i$.
In the same way, fragmentations occur from a gluon as shown in the figure.
Since there are two processes in $O(g^2)$ with a soft gluon
radiation, the second moment for the gluon is expected to
be larger than the others. These considerations lead to
the relation $M_u<M_s \lesssim M_g$ in Table \ref{tab:f0-config}.
Such a naive estimation should be a crude one, but it has been
shown to work for the moments of the pion, kaon,
and proton \cite{hkns07}, so that it is also expected to
be a reasonable guideline in other hadrons.

Next, functional forms are discussed in the $s\bar s$ picture. 
More energy is transferred to $f_0$ from the initial $s$
in the $O(g^2)$ process than the one from the initial $u$
due to an extra gluon emission in Fig. \ref{fig:ssbar}. It means
that the fragmentation function $D_s^{f_0}(z)$ is distributed
in the larger $z$ region in comparison with $D_u^{f_0}(z)$
because the $f_0$ energy is directly proportional to $z$.
Namely, they should have different functional forms and
their peak positions are different at small $Q^2$ ($\sim$1 GeV$^2$).
We express this situation as $z_u^{\rm max}<z_s^{\rm max}$
in Table \ref{tab:f0-config}. The form of the gluon fragmentation
function may not be simply compared with the quark processes.

In the same way, the second moments and functional forms
are roughly estimated for the tetraquark picture.
Since the fragmentations from $u$ and $s$ quarks
are equally favored processes in this case, their moments and
functions forms should be almost the same. The fragmentations into
$f_0$ proceed by creating $u\bar u$ (or $d\bar d$) and $s\bar s$
pairs as shown in Fig. \ref{fig:tetraquark}.
There are more fragmentation processes from a gluon, so
that the gluon moment is expected to be larger than the others.
In this way, we obtain the relations, $M_u \sim M_s \lesssim M_g$
and $z_u^{\rm max} \sim z_s^{\rm max}$, in Table \ref{tab:f0-config}.
Since the flavor composition of $f_0$ is simply considered in the above 
discussions, this relation could be also applied to the $K\bar K$ case.
However, the $K\bar K$ is a loose and extended bound state
so that its production probability in the fragmentation
is expected to be much smaller than that of the tetraquark state.

Although the nonstrange-$q\bar q$ and glueball configurations
seem to be unlikely according to recent theoretical investigations,
we also estimated possible relations in Table \ref{tab:f0-config}.
Since the estimation method is essentially the same, derivations are
not explained here. If $f_0$ were a nonstrange-$q\bar q$
meson, the relations $M_s<M_u<M_g$ and $z_u^{\rm max}>z_s^{\rm max}$ 
are expected, whereas they are 
$M_u \sim M_s < M_g$ and $z_u^{\rm max} \sim z_s^{\rm max}$
if it were a glueball.


The fragmentation functions are determined by a global analysis
of hadron-production data in $e^+ e^-$ annihilation \cite{ffs}.
There is recent progress on their analysis.
Uncertainties of the fragmentation functions are determined in
Ref. \cite{hkns07}, and it was shown that the gluon and
light-quark functions have large uncertainties for the pion,
kaon, and proton. Then, a global analysis with data in lepton
scattering and proton-proton collisions was also reported \cite{dss}.
This kind of global analysis is suitable for finding exotic hadrons
by noting the typical features in the favored and disfavored functions. 

All the possible configurations for $f_0$ indicate that
up- and down-quark compositions are the same; however, they are
generally different from the strange-quark and other ones.
Therefore, a natural and model-independent parametrization is
\begin{align}
D_{u}^{f_0} (z,Q_0^2) & = D_{\bar u}^{f_0} (z,Q_0^2)
= D_{d}^{f_0} (z,Q_0^2) =  D_{\bar d}^{f_0} (z,Q_0^2)
\nonumber \\
           &  = N_{u}^{f_0} z^{\alpha_{u}^{f_0}}
                (1-z)^{\beta_{u}^{f_0}},
\nonumber \\
D_{s}^{f_0} (z,Q_0^2) & = D_{\bar s}^{f_0} (z,Q_0^2)  
              = N_{s}^{f_0} z^{\alpha_{s}^{f_0}}
                (1-z)^{\beta_{s}^{f_0}},
\nonumber \\
D_{g}^{f_0} (z,Q_0^2) & 
              = N_{g}^{f_0} z^{\alpha_{g}^{f_0}}
                (1-z)^{\beta_{g}^{f_0}},
\\                
D_{c}^{f_0} (z,m_c^2) & = D_{\bar c}^{f_0} (z,m_c^2)
 = N_{c}^{f_0} z^{\alpha_{c}^{f_0}} (1-z)^{\beta_{c}^{f_0}} ,
\nonumber \\
D_{b}^{f_0} (z,m_b^2) & = D_{\bar b}^{f_0} (z,m_b^2)
 = N_{b}^{f_0} z^{\alpha_{b}^{f_0}} (1-z)^{\beta_{b}^{f_0}} ,
\nonumber            
\end{align}
where $N_i$, $\alpha_i$, and $\beta_i$ are the parameters to be
determined by a $\chi^2$ analysis of the data for
$e^++e^- \rightarrow f_0+X$ \cite{e+e-f0-data}. 
The initial scale is taken $Q_0^2$=1 GeV$^2$, and the masses
are $m_c$=1.43 GeV and $m_b$=4.3 GeV.
The details of the analysis method in the next-to-leading order
are explained in Ref. \cite{hkns07}.
Uncertainties of the determined functions are estimated by
the Hessian method \cite{hkns07}, which has been used also
in the studies of various parton distribution functions
\cite{unpol-error,errors}:
\begin{equation}
[\delta D_i^{f_0} (z)]^2=\Delta \chi^2 \sum_{j,k}
\left( \frac{\partial D_i^{f_0} (z,\xi)}{\partial \xi_j} \right)_{\hat\xi}
H_{jk}^{-1}
\left( \frac{\partial D_i^{f_0} (z,\xi)}{\partial \xi_k} \right)_{\hat\xi}
.
\label{eqn:ddih}
\end{equation}
Here, $\delta D_i^{f_0} (z)$ is the uncertainty of the fragmentation
function $D_i^{f_0} (z)$, $\Delta \chi^2 $ value is taken so that
the confidence level $P$ becomes the one-$\sigma$-error range ($P=0.6826$)
by assuming the normal distribution in the multi-parameter space,
$H_{ij}$ is the Hessian matrix, $\xi_i$ is a parameter, and
$\hat\xi$ indicates the optimum parameter set.

\begin{figure}[t]
\includegraphics[width=0.34\textwidth]{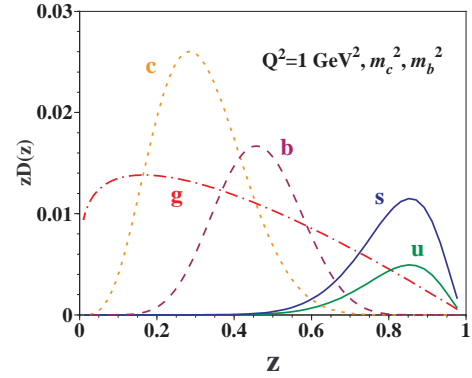}
\vspace{-0.3cm}
\caption{(Color online) Determined fragmentation functions
         of $f_0 (980)$ by the global analysis.
         The functions $zD_u^{f_0}$, $zD_s^{f_0}$, and $zD_g^{f_0}$
         are shown at $Q^2$=1 GeV$^2$, and
         the functions $zD_c^{f_0}$ and $zD_b^{f_0}$ are
         at $Q^2=m_c^2$ and $m_b^2$, respectively.}
\label{fig:f0-ffs}
\end{figure}

The number of $f_0$ data is very limited at this stage. In fact,
available data are merely twenty three. This situation makes
the analysis difficult in obtaining the minimum $\chi^2$ point. 
There are irrelevant parameters which do not affect the total $\chi^2$.
We decided to fix three parameters at $\beta_g=1$, $\alpha_u=10$,
and $\alpha_s=10$ because of the lack of data.
Then, the total number of parameters becomes twelve.
The minimum $\chi^2$ is obtained $\chi^2$/d.o.f.=0.907 in our analysis. 

The determined functions are shown in Fig. \ref{fig:f0-ffs}.
It is interesting to find that the up- and strange-quark functions are
distributed relatively at large $z$, and both functions have similar
shapes, whereas the gluon, charm-, and bottom-quark functions are
distributed at smaller $z$. It may indicate that both functions, 
$D_u^{f_0}$ and $D_s^{f_0}$, are equally favored ones, which implies
that the up-quark (and down-quark) is one of main components of
$f_0$ as well as the strange-quark. Furthermore, they are peaked
almost at the same points of $z$ ($z_u^{\rm max} \sim z_s^{\rm max}$),
which may be also considered as an evidence for the tetraquark structure
according to Table \ref{tab:f0-config}. However, if it is judged
from their second moment ratio ($M_u/M_s=0.43$), it looks like
the $s\bar s$ configuration. 

\begin{figure}[t]
\includegraphics[width=0.34\textwidth]{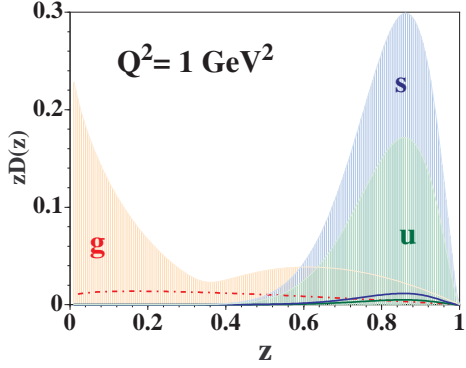}
\vspace{-0.3cm}
\caption{(Color online) Fragmentation functions,  
    $zD_u^{f_0}$, $zD_s^{f_0}$, and $zD_g^{f_0}$, and 
    their uncertainties are shown at $Q^2$=1 GeV$^2$.
    The uncertainties are shown by the shaded bands.}
\label{fig:f0-ffs-error}
\end{figure}

This conflict is mainly caused by the inaccurate determination of
the fragmentation functions although it may be understood by
admixture of the $s\bar s$ and tetraquark configurations.
In Fig. \ref{fig:f0-ffs-error},
the uncertainties of $zD_u^{f_0}$, $zD_s^{f_0}$, and $zD_g^{f_0}$
are shown at $Q^2$=1 GeV$^2$ together with the functions themselves.
We notice huge uncertainties which are an order of magnitude larger
than the determined functions. If their moments are calculated,
they have large errors:
$ M_u = 0.0012 \pm 0.0107$,
$ M_s = 0.0027 \pm 0.0183$, and
$ M_g = 0.0090 \pm 0.0046$.
From these results, the error of the moment ratio is estimated as
$M_u/M_s=0.43\pm 6.73$, which makes it impossible to discuss
the effect of the order of 50\%. In this way, we find the structure
of $f_0$ cannot be determined by the current $e^+e^-$ data.

It is the purpose of this work to point out that structure of exotic
hadrons should be determined by the fragmentation functions. 
Accurate measurements of hadron-production cross sections can be
used for determining their internal quark and gluon configurations
as explained in this paper by taking $f_0(980)$ as an example. 
We have shown that $s\bar s$ and tetraquark configurations,
and also nonstrange-$q\bar q$ and glueball states, should
be distinguished by the second moments and functional forms
of the favored and disfavored functions. Especially, the ratio
of the $u$-quark moment to the $s$-quark one should be useful
to judge the configuration.

In order to determine the internal structure, the flavor separation
is important especially because the difference between the up- and
strange-quark functions is the key to find the structure of $f_0$.
First, charm- and bottom-quark tagged data should be provided for
$f_0$ as they have been obtained for the pion, kaon, and proton.
Then, the charm- and bottom-quark functions should be determined
accurately. Second, semi-inclusive $f_0$-production data in lepton-proton
scattering can be used for distinguishing between up- and strange-quark
fragmentations because the initial quark distributions are different
in the proton. These flavor separations will become possible by
future experimental analyses. Our work is a starting point for exotic
hadron search by suggesting the relations in the second moments and
the functional forms and by indicating the current experimental situation
as the uncertainty bands.

The fragmentation functions of $f_0$ and their uncertainties have
been determined by the global analysis of $f_0$ production data.
At this stage, the $e^+e^-$ data are not precise enough;
however, accurate experimental measurements could create
a field of exotic hadrons which are beyond the naive $q\bar q$
and $qqq$ type ones. Currently, analyses are in progress
by the Belle collaboration \cite{belle-ffs} to provide accurate 
fragmentation functions. They are especially important because
the functions are measured at small $Q^2$ ($\ll M_Z^2$), so that
scaling violation can be investigated to find the gluon functions 
\cite{hkns07}. It is also important to have accurate measurements
for ordinal mesons such as $\phi (1020)$ and $f_2 (1270)$ in order to establish
the $f_0$ configuration by comparing their favored and disfavored
functions with the ones of $f_0$. We could investigate other exotic
hadrons in the same way by their fragmentation functions.



\end{document}